\documentclass{emulateapj-rtx4}
\shorttitle{FCC046:~a candidate polar-ring dE} \shortauthors{De Rijcke
  et al.}

\begin{document}

\title{FCC046:~a candidate gaseous
  polar ring dwarf elliptical galaxy in the Fornax Cluster }

\author{S. De Rijcke\altaffilmark{1}, P. Buyle\altaffilmark{1}, M. Koleva\altaffilmark{1}}

\altaffiltext{1}{Ghent University, Dept. Physics \& Astronomy,
  Krijgslaan 281, S9, B-9000, Ghent, Belgium}

\begin{abstract}
FCC046 is a Fornax Cluster dwarf elliptical galaxy. Optical
observations have shown that this galaxy, besides an old and
metal-poor stellar population, also contains a very young centrally
concentrated population and is actively forming stars, albeit at a
very low level. 

Here, we report on 21~cm observations of FCC046 with the Australia
Telescope Compact Array (ATCA) which we conducted in the course of a
small survey of Fornax Cluster early-type dwarf galaxies. We have
discovered a $\sim 10^7$~M$_\odot$ H{\sc i} cloud surrounding FCC046.
We show that the presence of this significant gas reservoir offers a
concise explanation for this galaxy's optical morphological and
kinematical properties.

Surprisingly, the H{\sc i} gas, as evidenced by its morphology and its
rotational motion around the galaxy's optical major axis, is
kinematically decoupled from the galaxy's stellar body. This is the
first time such a ring of gaseous material in minor-axis rotation is
discovered around a dwarf galaxy.
\end{abstract}


\section{Introduction} \label{sect:exdat}

FCC046 was included in the Fornax Cluster Catalog by \citet{fe89}. It
was classified as a dE4, i.e. as a rather flattened dwarf elliptical
galaxy, with a total B-band apparent magnitude $m_B=15.99$~mag and a
half-light radius of $R_e = 6.7''$. We adopt a distance of 20.3~Mpc
\citep{je03} for the distance to the Fornax Cluster, yielding a total
absolute magnitude of $M_B=-15.55$~mag and a half-light radius of $R_e
= 660$~parsec for FCC046 \citep{de05}. Using VLT photometry and
spectroscopy, \citet{dd04} showed that FCC046, despite its pronounced
flattening, has zero net stellar rotation about neither the projected
major or the minor axis. It also has a very pronounced nucleus which
is resolved from the ground. The nucleus' most striking feature is
that, surprisingly, it is off-center with respect to the galaxy's
outer isophotes by almost 1~arcsec. This displacement has been
interpreted as a consequence of the counter-rotation instability.

\section{Existing optical observations}

Observations with the FLAIR-II spectrograph on the UK
Schmidt Telescope by \citet{dr01} have shown that FCC046 has H$\alpha$
emission, with an equivalent width of 2.1~{\AA} over a fibre
6.7~arcsec accross centered on the galaxy. The presence of ionized gas was
interpreted as evidence for star formation, casting doubt on its
classification as a ``true'' dwarf elliptical.

This has later been confirmed by \citet{de03}, who showed, based on
VLT photometry, that FCC046 has strong positive color gradients and
that its nucleus is surrounded by 6 sources of H$\alpha$ emission. The
largest of the emission sources have properties (diameters and
H$\alpha$+[N{\sc ii}] luminosities) similar to those of supernova
remnants. The smaller ones could be H{\sc ii} regions or nebulae
around Wolf-Rayet stars. Blindly applying the heuristic relation
between the star-formation rate and the H$\alpha$ emission line's
equivalent width \citep{ke92} leads to a star-formation rate of $\sim
10^{-3}$~M$_\odot$/year. The colour gradient also suggests the
occurrence of centrally concentrated recent star formation in FCC046.

%% \begin{figure}
%% \epsscale{1.1}
%% \plotone{fcc046_age_feh.eps}
%% \caption{Radial profiles of SSP-equivalent age (top panel) and
%%   metallicity, quantified via [Fe/H] (bottom panel), in FCC046. The
%%   vertical dashed lines indicates the position of the half-light
%%   radius. Figure adopted from \citet{ko09}.
%% \label{fig:fcc046_age_feh.eps}}
%% \end{figure}

\begin{figure*}
\epsscale{1.1}
\plottwo{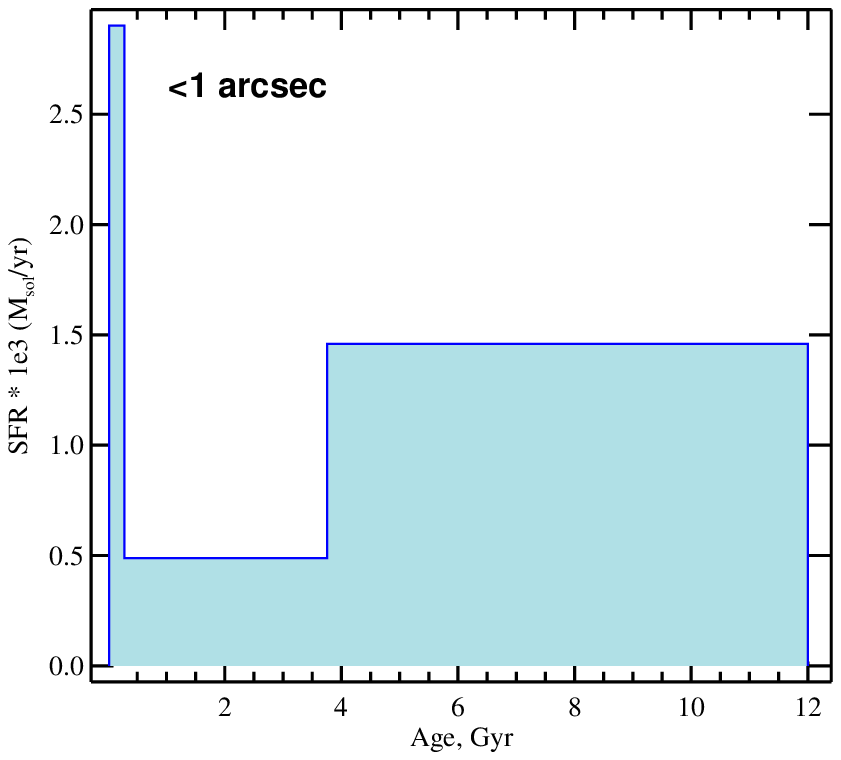}{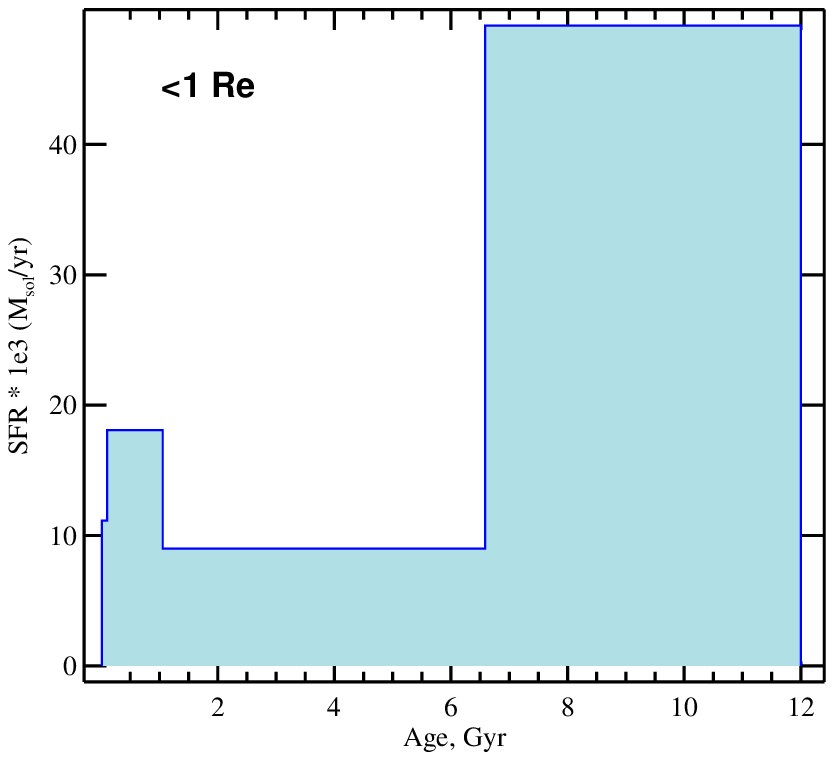}
\caption{Left panel:~the star-formation rate as a function of stellar
  population age as derived from the stacked spectra inside
  1~arcsec. Right panel:~the same but derived from stacked spectra
  between 1~arcsec and 1~$R_e$ (6.7~arcsec). This figure is based on a
  careful re-analysis of the data presented in \citet{ko09}.
\label{fig:SFH}}
\end{figure*}

Applying the full-spectrum fitting code ULySS to an optical VLT
spectrum of FCC046, \citet{ko09} showed that the SSP-equivalent age of
the stellar population increases from less than 1~Gyr near the center
to $\sim 5$~Gyr beyond 1~kpc (see fig. 2 of \cite{ko09}). At the same
time, the mean luminosity weighted metallicity, quantified by the
SSP-equivalent [Fe/H], increases very slightly from a central value of
[Fe/H]$\sim -1.0$~dex to [Fe/H]$\sim -0.8$~dex at 1~$R_e$.

By fitting a weighted sum of several SSPs to the spectra one can
reconstruct a star-formation history \citep{lb04}. The number of age
bins used in this fit was steadily increased until the goodness-of-fit
stopped increasing significantly. The bounds of the bins were varied
to ensure that the age and metallicity of each SSP were well within
their respective bounds (e.g. if the age of an SSP coincided with the
lower age bound of its bin, this bound was lowered).

In Table \ref{tab1}, we present the results from fits to spectra
summed within the inner seeing disk (1~arcsec accross) and within one
$R_e$ (excluding the inner seeing disk). Within the inner
1~arcsecond, young stars make up 59~\% of the light; within the inner
1~R$_e$, they constitute 47~\% of the light, with about 10~\% of
newborn stars. Apparently, star formation picked up about 300~Myr ago
in the body of the galaxy while the increase in the inner 1~arcsecond
happened only $\sim 70~$Myr ago. Clearly, the positive age gradient
observed in FCC046 is caused by the presence of centralised ongoing
star formation, resulting in a very young stellar population whose
fractional light contribution decreases with radius. 

Dividing the stellar mass in each age bin by its age width yields the
mean star-formation rate in each bin, shown in Fig. \ref{fig:SFH}. The
central 1~arcsecond wide disk has a B-band luminosity of $L_B(1'')=1.0
\times 10^7~L_{B,\odot}$; the inner 1~R$_e$ (excluding the inner disk)
has a luminosity $L_B(1~R_e)=1.2 \times 10^8~L_{B,\odot}$. The
corresponding masses for the stellar populations inside the inner 1
arcsec and inside 1~$R_e$, derived with the Pegase.HR evolutionary
code for a Salpeter IMF using the SFH reconstruction of ULySS, are
M$(1'')=1.5 \times 10^7$~M$_{\odot}$ and M$(1~R_e)=3.3 \times
10^8$~M$_{\odot}$. We normalize the SFR such that the time-integral
over the star-formation history in each radial bin yields the correct
total mass. The recent star-formation event appears to have been very
pronounced in the inner regions. There, the star-formation rate
increased by over a factor of 5 compared with the mean rate over the
previous $\sim 4$~Gyr, although the time resolution of the SFH does
not allow us to pinpoint other similar bursts older than $\sim 1$~Gyr.

The oldest stars (age $>3$~Gyr) have metallicities in the range
[Fe/H]$\sim -2$ to $-1$ throughout the whole galaxy. Judging from
Table \ref{tab1}, the young and intermediate-age stellar populations
have metallicities consistent with solar while the 12~Gyr old
population has a higher metallicity in the outer radial bin
([Fe/H]$\sim -1.5$ versus [Fe/H]$\sim -2.3$ inside the inner
1~arcsec). This appears to be the cause for the slight radial increase
of the SSP-equivalent metallicity. The metallicity of the very
youngest stellar populations is rather uncertain since their massive
stars, which dominate their optical light output, have no conspicuous
metallicity-sensitive absorption features. However, there appears to
be a tendency for the very youngest stellar populations to have
metalicities below solar.

\begin{table}
\caption{Stellar populations inside the inner 1 arcsecond (top) and
  inside 1~$R_e$ (bottom) \label{tab1}}
\begin{tabular}{|c|c|c|}\hline
\multicolumn{3}{|l|}{Inside 1~arcsecond} \\ \cline{1-3}
Age & light fraction & [Fe/H] \\ \cline{1-3}
$67 \pm 5$~Myr & 59\% & $-0.11 \pm 0.07$ \\
$1176 \pm 352$~Myr & 14\% & $-0.01 \pm 0.27$ \\
$12000$~Myr (fixed) & 27\% & $-2.27 \pm 0.13$ \\ \cline{1-3}
\multicolumn{3}{|l|}{Inside 1~$R_e$} \\ \cline{1-3}
Age & light fraction & [Fe/H] \\ \cline{1-3}
$34 \pm 10$~Myr & 11\% & $-0.80 \pm 0.45$ \\
$309 \pm 51$~Myr & 36\% & $0.16 \pm 0.12$ \\
$3613 \pm 1857$~Myr & 8\% & $0.27 \pm 0.38$ \\
$12000$~Myr (fixed) & 45\% & $-1.46 \pm 0.16$ \\ \cline{1-3}
\end{tabular}
\end{table}
Within the inner 1~arcsecond, the nucleus contributes $\sim 80$~\% of
the light while $\sim 20$~\% comes from the galaxy's main body,
adopting the latter's S\'ersic profile parameters from
\citet{ko09}. If we assume that the light fraction of 11~\%
contributed by very young stars is typical for the galaxy's body (see
Table \ref{tab1}) then the 59~\% of light from the central 1~arcsecond
coming from very young stars, with ages below 100~Myr, translates into
a nucleus whose youngest stellar populations contribute $\sim 70$~\%
of its light. The mass of the nucleus is $M_{\sf nuc} \approx 8 \times
10^6$~M$_\odot$, 7~\% of which consists of very young stars. For this
estimate, we adopted a $M/L \approx 0.1$ in solar units for a 70~Myr
old population from PEGASE-HR \citep{lb04}. Therefore, most of the
mass of the nucleus was in place before the most recent star-formation
event.

With such a massive nucleus, box orbits are scattered onto tube and
chaotic orbits; conservation of angular momentum then requires that
direct and retrograde loop orbits be equally populated
\citep{ec94,mq98,va10}. If this scenario holds, FCC046 would have
developed an axially symmetric counter-rotating stellar body, prone to
the counter-rotation instability, over the course of only a few
crossing times \citep{dd04}.

\section{New H{\sc i} observations} \label{sect:HI}

\begin{figure*}
\epsscale{1.0}
\plotone{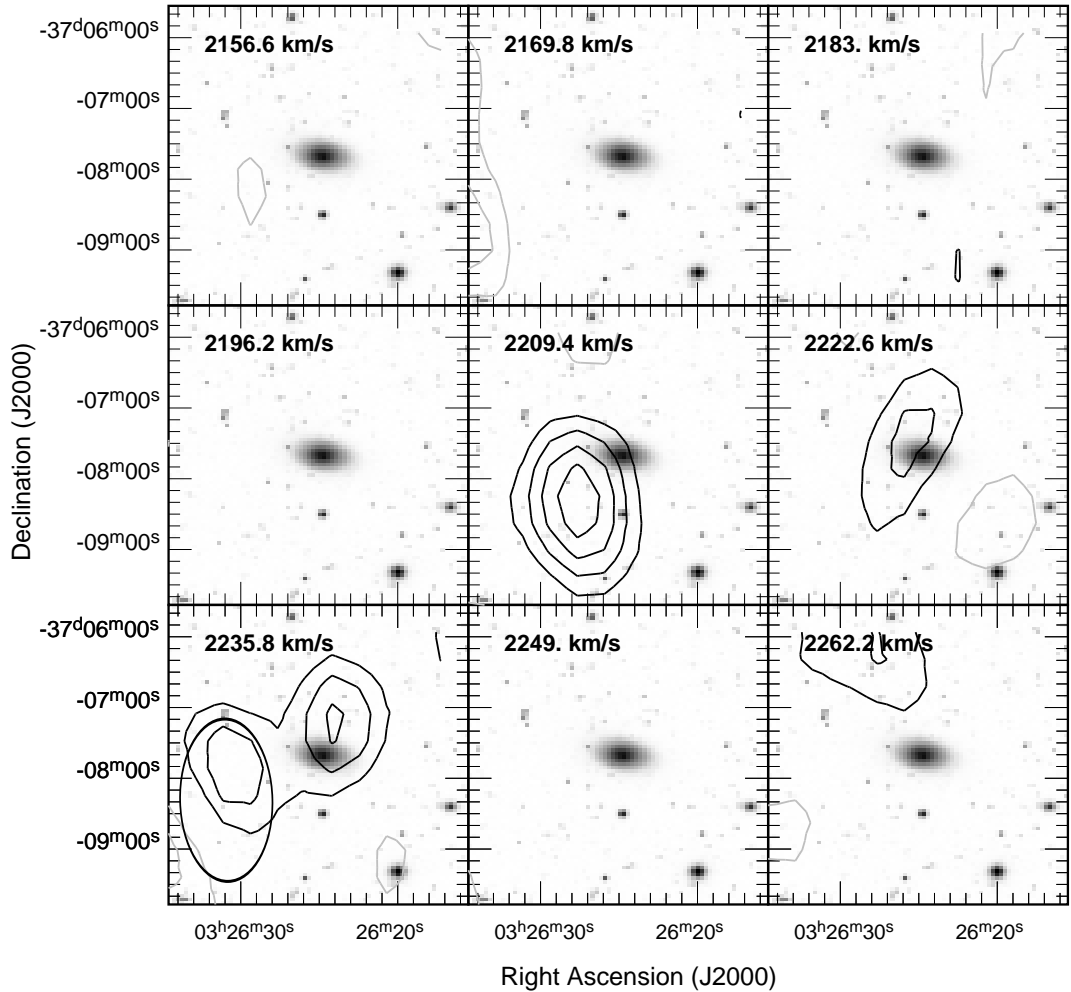}
\caption{13.2~km/s wide channel maps of FCC046. The beam size is
  indicated by the ellipse in the bottom left panel. Contours are
  drawn for flux levels of $\pm 2 \sigma$, $\pm 3 \sigma$, {\ldots}
  with $1\sigma=3$~mJy/beam.  Grey contours represent negative flux
  values. The VLT/FORS2 B-band image in the background is taken from
  \citet{de03}.
\label{fig:velocity.eps}}
\end{figure*}

We used the Australia Telescope Compact Array (ATCA) during February
2006 to search for 21~cm emission in a small sample of Fornax Cluster
dEs. The observations took place during daytime but were not disturbed
by solar radio-frequency interference. The ATCA was positioned in the
large EW367 configuration with baselines ranging from 46~m up to
4408~m. Due to the typical small velocity widths of dwarf galaxies, we
opted for a correlator setup with 256 channels of width 31.25 kHz,
giving a total baseband of 8 MHz or roughly 1700 km/s at the distance
of FCC046. No on-line Hanning smoothing was applied, resulting in a
velocity resolution of 6.6~km/s. The observation was initiated with a
calibration on the source 1934-638 which acted as primary
calibrator. After the initial calibration, this source was observed
for 15 minutes, followed by alternating observations of the phase
calibrator 0332-403 (integration of 5 minutes) and of the targeted
dwarf galaxy (integration of 40 minutes). The total integration time
on each target (including calibration) was 12h.

The standard data reduction steps (phase, amplitude and bandpass
calibration) were performed with the MIRIAD package \citep{sa95}, the
standard ATCA data analysis program. We subtracted the continuum by
performing a first-order fit to the visibilities over the line-free
channels that were not affected by the edge effects of the band (300
km/s on each edge). The data cubes were created by using natural
weighting. A minimal cleaning of 500 iterations was performed. Our
final data cube had a synthesized beam of $147 \times 78$~arcsec$^2$
and a noise of 3 mJy/beam. These steps resulted in a detection of
FCC046 in at least 3 subsequent 6.6~km/s channels in the centre of our
baseband at a velocity of 2209 km/s, very close to the optical
velocity of FCC046 at 2220 km/s. Channel maps of FCC046, rebinned to
13.2~km/s wide channels, are presented in Fig. \ref{fig:velocity.eps}.

\begin{figure}
\epsscale{1.15}
\plotone{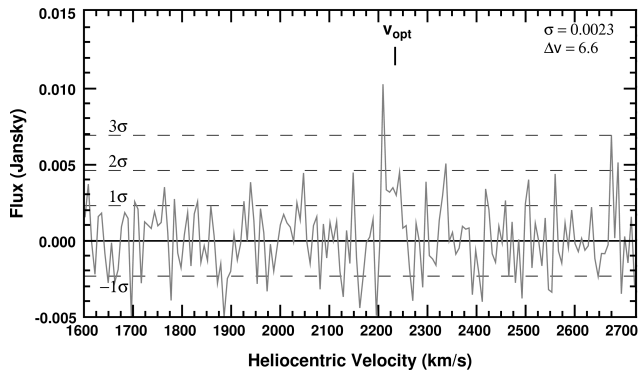}
\caption{Base-line subtracted H{\sc i} spectrum of FCC046, extracted
  from a $4 \times 4$~arcmin$^2$ box centred on the central radio
  position of FCC046. The galaxy's optical velocity is indicated as
  $v_{\rm opt}$. 
\label{fig:histo.eps}}
\end{figure}

To derive a spectrum we summed the flux within a $4 \times
4$~arcmin$^2$ box centred on the central radio position of FCC046,
which was derived from the total H{\sc i} intensity map (see
Fig. \ref{fig:histo.eps}). The global H{\sc i} profile shows a double horned spectrum
suggesting rotation. A single Gaussian was fitted to the H{\sc i}
profile, giving a velocity linewidth of 52 km/s at the 20 per cent
level and 34 km/s at the 50 per cent level. After correction for
broadening and random motions (method of \citet{vs01}) we find
$W_{20}=43$~km/s and $W_{50}=32$~km/s.

After summing the channels that contain emission of the galaxy, we
found a total velocity-integrated H{\sc i} flux density of 0.15 Jy
km/s, resulting in a total estimated H{\sc i} mass of $1.45 \times
10^7$~M$_\odot$.

\section{Discussion} \label{sect:disc}

\begin{figure*}
\epsscale{1.15}
\plottwo{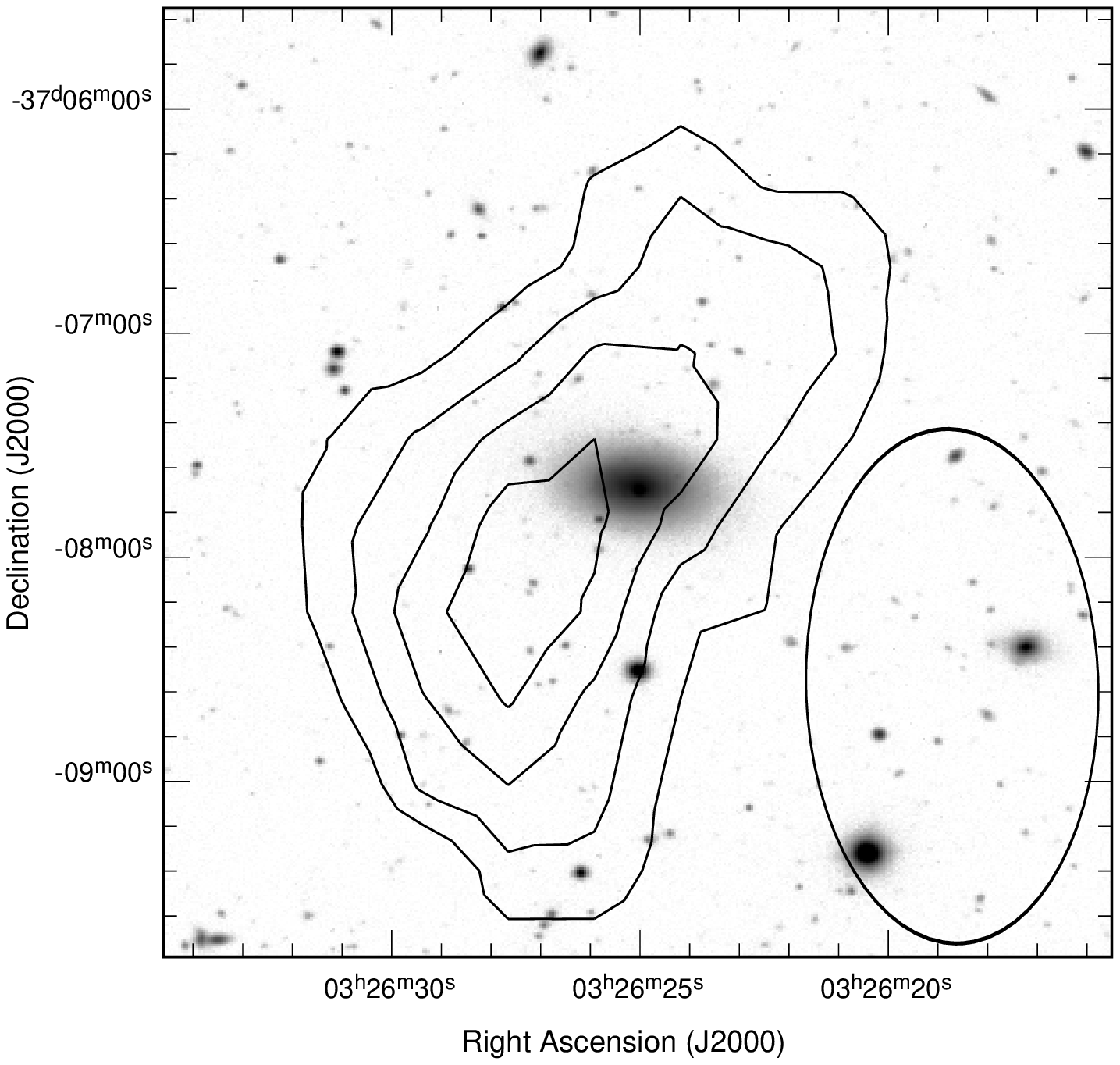}{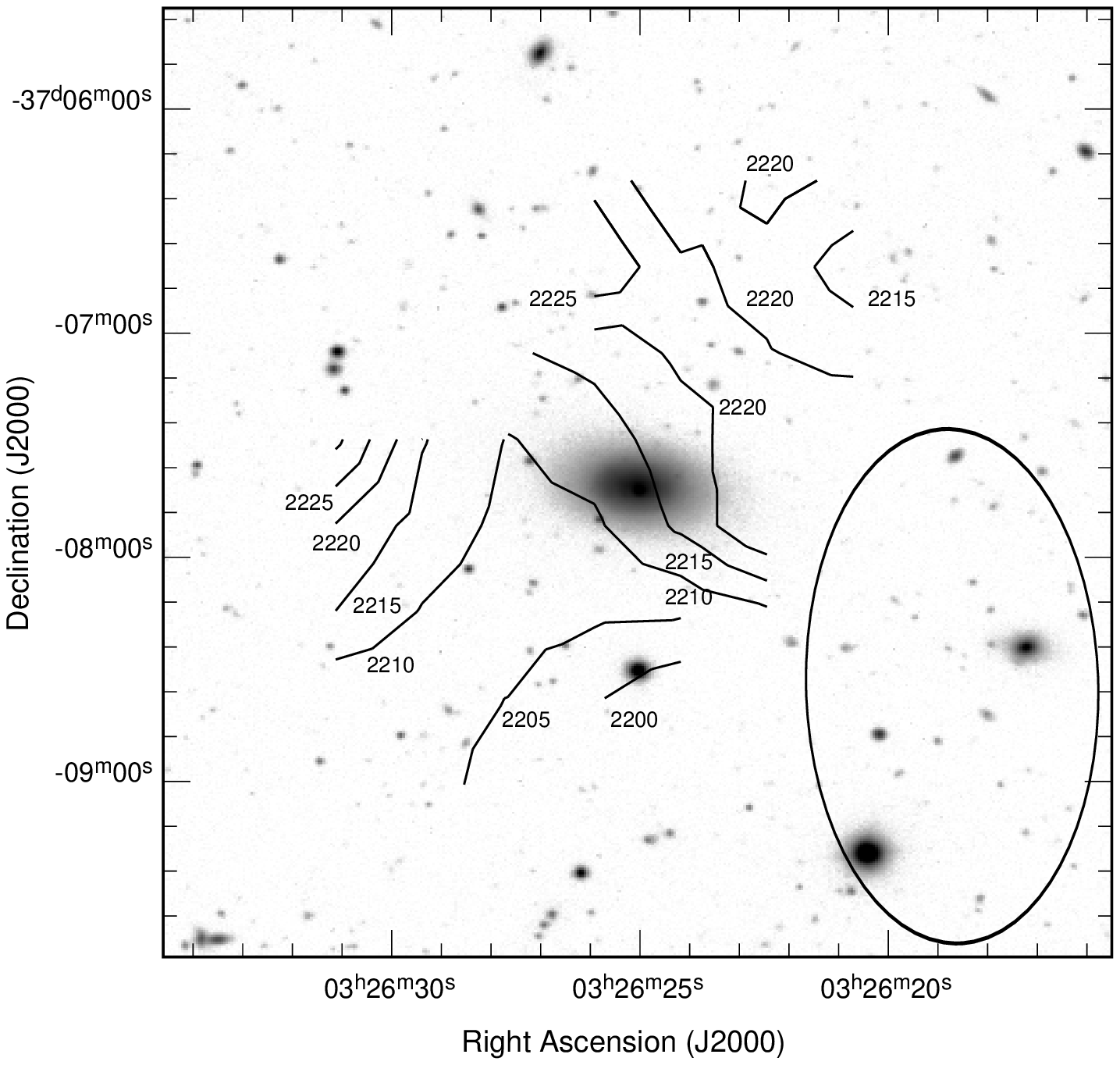}
\caption{Zeroth-order (left) and first-order (right) moment maps of
  the 21~cm emission of FCC046. The size and shape of the synthesized
  beam is indicated in the bottom right corner of each panel. While
  the beam size is quite significant, the emission is clearly
  elongated in the direction perpendicular to the galaxy's stellar
  body. There is also a hint of a velocity progression from the
  galaxy's southern side towards the north-west. Zeroth-order moment
  contours are given for the $1\sigma$, $2\sigma$, $3\sigma$, and
  $4\sigma$ levels with $1\sigma=0.06$~Jansky~km~s$^{-1}$~beam$^{-1}$.
  The first-order moment map is shown only above the $2\sigma$ level.
\label{fig:moment0.eps}}
\end{figure*}

The most striking feature of the 21~cm emission of FCC046 is that it
is elongated in a direction almost perpendicular to the galaxy's
stellar body. This is shown clearly in Fig. \ref{fig:moment0.eps} in
which a zeroth-order map of the 21~cm emission of FCC046 is
presented. While the beam size is, admittedly, quite significant, the
emission region is clearly resolved and, at FWHM, is almost 1.5 times
as large as the beam in the north-south direction. Its east-west
extent is not resolved. The first-order moment map shows evidence for
a velocity progression from the galaxy's southern side towards the
north-west (the velocity increase eastwards of the galaxy is due to a
$2\sigma$ peak in a single channel and is likely to be spurious). If,
allowing for the noise level and resolution of our observations, we
interpret this as ordered rotation this means that the H{\sc i} gas
surrounding FCC046 is rotating essentially around the optical galaxy's
major axis.

Polar rings, i.e. rings of stars and/or gas and dust orbiting in a
plane perpendicular to their host galaxy's equatorial plane, are known
in a small fraction of low-mass early-type galaxies
\citep{whi90,co06}. Cosmological simulations \citep{mac06} suggest
that these galaxies acquire this material through cold accretion of
gas filaments. A similar scenario could apply to FCC046. The position
of FCC046, at $3.5^\circ$ from the cluster center \citep{bu05}, places
it far outside the X-ray halo of the Fornax Cluster, which does not
extend significantly beyond the NGC1399/NGC1404/NGC1387 central region
\citep{jo97}. This means that any gas accreted onto FCC046 will not
immediately be removed by ram-pressure stripping, keeping it available
for star formation. It then stands to reason that FCC046, an
``ordinary'' dE with a stellar population with a mean age over $\sim
3$~Gyr, has accreted a few $10^7$~M$_\odot$ of gas orbiting roughly
perpendicularly to its stellar body. The slightly distorted morphology
of the gas may be due to the orbital movement of FCC046 through the
Fornax cluster, combined with its weak gravitational field, or by
warping. Moreover, if FCC046 has a strongly oblate or prolate
gravitational potential, the polar ring gas is expected to rapidly
flow inwards where it can be used for star formation \citep{ch92}.
This inflow, flooding the ring's central hole with gas, would occur on
a timescale corresponding to only a few rotation periods. In the case
of FCC046, this would amount to a few 100~Myrs which is in line with
the observed upturn of the star-formation rate.

Assuming we are indeed observing a gas ring, the observed velocity
gradient will be much lower than what could be expected from the
galaxy's rotation curve. In the central hole, it is precisely the gas
on orbits tangent to the line of sight that is missing. The gas within
the ring (back and front side) has much less favorable projections of
its rotation velocity onto the line of sight and therefore results in
small projected velocities. Only along sufficiently distant lines of
sight, those that avoid the central hole, is it possible to observe
the full rotation of the gas. However, given the theoretically
predicted ongoing warping and disruption of the ring, even there gas
is most likely not moving at the local rotation velocity. Along with
the severe beam smearing, this helps explains the observed small
velocity gradient in the right panel of Fig. \ref{fig:moment0.eps}.

Clearly, the hypothesis that FCC046 has accreted a fresh reservoir of
gas that triggered a recent star-formation event offers a concise
explanation for most of this galaxy's observed properties. Other
explanations work far less well. For instance, the observed H{\sc i}
gas could have been expelled by the centralized star-formation event,
rendering it an outflow rather than an infall. One might expect that,
in a flattened dwarf galaxy, such outflows are preferably aligned
along the minor axis, as is sometimes observed \cite{mf99}. However,
the recent star-formation rate measured in FCC046 is orders of
magnitudes smaller than what is observed in e.g. the star-bursting
bipolar outflow dwarf M82, which has a central star-formation rate
upwards of $1$~M$_\odot$/year \citep{fs03}. This makes it unlikely
that FCC046 could produce the $G{\rm M}_{\rm gal}{\rm M}_{\rm
  gas}/{\rm 1~kpc}\sim 10^{54}$~erg, the equivalent of over $\sim
1000$ combined supernova explosions \citep{cl12}, needed to lift
M$_{\rm gas}= 1.45 \times 10^7$~M$_\odot$ of gas out to a distance of
1~kpc in a M$_{\rm gal} \sim 10^9$~M$_\odot$ galaxy. Moreover,
realistic simulations of flattened dwarf galaxies fail to produce such
collimated outflows if the star-formation activity is not very
centrally concentrated \citep{sc11}. Another possibility is that we
are seeing gas being removed from FCC046 by ram-pressure stripping
\citep{mb00,ma03,ma06}. However, FCC046 is very far outside the X-ray
halo of the Fornax Cluster, making the stripping scenario highly
unlikely.

%% This would be the second example of minor-axis rotation to be
%% observed in a dwarf galaxy, after And~{\sc ii} in which stellar
%% minor-axis rotation was discovered \citep{ho12}.

\section{Conclusions} \label{sect:conc}

Based on optical observations, FCC046, a dwarf galaxy in the Fornax
Cluster, has properties that make it stand out from the general dwarf
elliptical galaxy population. All evidence supports the conclusion
that $\sim 300$~Myr ago, the star-formation rate in FCC046 increased
dramatically. The strength of this recent star-formation event
decreases with radius since these young stars contribue $\sim 60$~\%
of the light within the inner arcsecond (a region largely coinciding
with the galaxy's nucleus) while they make up only 47~\% of the light
within one half-light radius, excluding the central arcsecond. Another
odd feature is this galaxy's relatively massive, off-center
nucleus. Given the fact that FCC046, despite its significant
flattening, has zero net rotation it is possible that the well-known
counter-rotation instability is responsible for driving the nucleus
off-center.

In this paper, we have presented H{\sc i} observations with the ATCA
of FCC046 that may offer a concise explanation for its optical
properties. We have discovered a $\sim 10^7$~M$_\odot$ H{\sc i} cloud
surrounding FCC046 which, as evidenced by its morphology and its
rotational motion around the galaxy's optical major axis, is
kinematically decoupled from FCC046's stellar body. It seems plausible
that this gas reservoir has been accreted by FCC046 on a highly
inclined orbit and, as it flows to the center of the gravitational
well, fuels centrally concentrated star formation.


\begin{thebibliography}{}
\bibitem[Buyle et al.(2005)]{bu05} Buyle, P., De Rijcke, S.,
  Michielsen, D., Baes, M., Dejonghe, H., 2005, MNRAS, 360, 853-858
\bibitem[Christodoulou et al.(1992)]{ch92} Christodoulou D. M., Katz
  N., Rix H.-W., Habe A., 1992, ApJ, 395, 113-118
\bibitem[Cloet-Osselaer et al.(2012)]{cl12} Cloet-Osselaer A., De
  Rijcke S., Schroyen J., Dury V., 2012, MNRAS, 423, 735-745
\bibitem[Cox et al.(2006)]{co06} Cox, A. L., Sparke, L. S., van
  Moorsel, G., 2006, AJ, 131, 828-836
\bibitem[De Rijcke et al.(2003)]{de03} De Rijcke, S., Zeilinger,
  W. W., Dejonghe, H., Hau, G. K. T., 2003, MNRAS, 339, 225-234
\bibitem[De Rijcke et al.(2005)]{de05} De Rijcke, S., Michielsen, D.,
  Dejonghe, H., Zeilinger, W. W., Hau, G. K. T., 2005, A\&A, 438, 491-505
\bibitem[De Rijcke \& Debattista(2004)]{dd04} De Rijcke, S. \&
  Debattista, V. P., 2004, ApJl, 1, L25-L28
\bibitem[Drinkwater et al.(2001)]{dr01} Drinkwater, M. J., Gregg,
  M. D., Holman, B. A., Brown, M. J. I., 2001, MNRAS, 326, 1076-1094
\bibitem[Evans \& Collett(1994)]{ec94} Evans, N. W., \& Collett, J. L.,
  1994, ApJ, 420, L67-L70
\bibitem[Ferguson(1989)]{fe89} Ferguson, H. C., 1989, AJ, 98, 367-418
\bibitem[F\"orster Schreiber, Genzel, Lutz(2003)]{fs03} F\"orster
  Schreiber N. M., Genzel R., Lutz D., 2003, ApJ, 599, 193-217
\bibitem[Ho et al.(2012)]{ho12} Ho, N., Geha M., Munoz R. R.
  Guhathakurta P., Kalirai J., Gilbert K. M., Tollerud E.,
  Bullock J., Beaton R. L., Majewski S. R., 2012, ApJ, 758, 124-135
\bibitem[Jerjen(2003)]{je03} Jerjen, H., A\&A, 2003, 398, 63-79
\bibitem[Jones et al.(1997)]{jo97} Jones, C., Stern, C., Forman, W.,
  Breen, J., David, L., Tucker, W., Franx, M., 1997, ApJ, 482, 143-155
\bibitem[Kennicutt(1992)]{ke92} Kennicutt R. C., 1992, ApJ, 388, 310-327
\bibitem[Koleva et al.(2009)]{ko09} Koleva, M., De Rijcke, S., Prugniel,
  Ph., Zeilinger, W. W., Michielsen, D., 2009, MNRAS, 396, 2133-2151
\bibitem[Le Borgne et al.(2004)]{lb04} Le Borgne D., Rocca-Volmerange
  B., Prugniel P., Lançon A., Fioc M., Soubiran C., 2004, A\&A, 425,
  881-897
\bibitem[Macci\`o et al.(2006)]{mac06} Macci\`o, A. V., Moore, B.,
  Stadel, J., 2006, ApJL, 626, L25-L28
\bibitem[Mac Low \& Ferrara(1999)]{mf99} Mac Low, M. \& Ferrara, A.,
  1999, ApJ, 513, 142-155
\bibitem[Marcolini, Brighenti, D'Ercole(2003)]{ma03} Marcolini,
  A., Brighenti, F., D'Ercole, A., 2003, MNRAS, 345, 1329-1339
\bibitem[Mayer et al.(2006)]{ma06} Mayer, L., Mastropietro, C., Wadsley,
  J., Stadel, J., Moore, B., 2006, MNRAS, 369, 1021-1038
\bibitem[Merritt \& Quinlan(1998)]{mq98} Merritt, D., \& Quinlan,
  G. D. 1998, ApJ, 498, 625
\bibitem[Mori \& Burkert(2000)]{mb00} Mori, M. \& Burkert, A., 2000,
  ApJ, 538, 559-568
\bibitem[Sault, Teuben, \& Wright(1995)]{sa95} Sault, R. J., Teuben,
  P. J., \& Wright, M. C. H. 1995, in ASP Conf. Ser. 77, Astronomical
  Data Analysis Software and Systems IV, ed. R. A. Shaw, H. E. Payne,
  \& J. J. E. Hayes (San Francisco: ASP), 433
\bibitem[Schroyen et al.(2011)]{sc11} Schroyen J., de Rijcke S.,
  Valcke S., Cloet-Osselaer A., Dejonghe H., 2011, MNRAS, 416, 601
\bibitem[Valluri et al.(2010)]{va10} Valluri, M., Debattista, V. P.,
  Quinn, T., Moore, B., 2010, MNRAS, 403, 525-544
\bibitem[Verheijen \& Sancisi(2001)]{vs01} Verheijen M. A. W. \&
  Sancisi R., 2001, A\&A, 370, 765
\bibitem[Whitmore et al.(1990)]{whi90} Whitmore B. C., Lucas
  R. A., McElroy D. B., Steiman-Cameron T. Y., Sackett
  P. D., Olling R P., 1990, AJ, 100, 1489-1522
\end{thebibliography}
\end{document}